\documentclass[10pt,conference]{IEEEtran} 
\IEEEoverridecommandlockouts
\usepackage{cite}
\usepackage{amsmath,amssymb,amsfonts}
\usepackage{algorithmicx}
\usepackage{graphicx}
\usepackage{textcomp}
\usepackage[dvipsnames]{xcolor}
\usepackage{amsmath}
\usepackage{marginnote}
\usepackage{url}
\usepackage{courier}
\usepackage[T1]{fontenc}
\usepackage{booktabs}
\usepackage{todonotes}
\usepackage{lipsum}
\usepackage{xargs}    
\usepackage{color,soul}
\usepackage{xspace}
\usepackage[ruled]{algorithm}
\usepackage[noend]{algpseudocode}
\DeclareMathOperator*{\argmax}{argmax}
\DeclareMathOperator*{\argmin}{argmin}

\graphicspath{ {./Figs/} }

\def\BibTeX{{\rm B\kern-.05em{\sc i\kern-.025em b}\kern-.08em
    T\kern-.1667em\lower.7ex\hbox{E}\kern-.125emX}}

\newcommand{\ie}{, i.e.,\xspace}
\newcommand{\eg}{, e.g.,\xspace}
\newcommand{\et}{et al.\xspace}

\newcommand{\z}[1]{\texttt{Znn.com}}

\newcommand{\ml}[1]{\emph{external}}
\newcommand{\al}[1]{\emph{internal}}     
\newcommand{\extr}{\textsc{HZ$_{e}$}\xspace}
\newcommand{\intr}{\textsc{HZ$_{i}$}\xspace}
\newcommand{\bl}[1]{\emph{baseline}} 
\newcommand{\app}{\textsc{HypeZon}\xspace}
\newcommand{\reqone}{\textbf{RQ1}\xspace}
\newcommand{\reqtwo}{\textbf{RQ2}\xspace}
\newcommand{\reqthree}{\textbf{RQ3}\xspace}

\newcommand{\chp}{\textsc{Chp$_{dtr}$}\xspace} 

\newcommand{\late}[1]{\texttt{latency}}
\newcommand{\budget}[1]{\texttt{overBudget}}
\newcommand{\utiliz}[1]{\texttt{underUtilized}}
\newcommand{\quality}[1]{\texttt{lowQuality}}

 \newcommand{\discharge}[0]{\texttt{Discharge-server}} 
 \newcommand{\enlist}[1]{\texttt{Enlist-server}}
 \newcommand{\increase} [1]{\texttt{Increase-quality}}
 \newcommand{\reduce}[1]{\texttt{Reduce-quality}}

\definecolor{Salmon}{rgb}{1,0.5,0}

\newcommand{\Note}[1]{\marginnote{\tiny}}
\newcommandx{\SG}[1]{\todo[textcolor=Sepia,linecolor=Melon!70,backgroundcolor=Dandelion!30,bordercolor=Melon!20,prepend,  size=\tiny]{SG:~#1}}

\newcommandx{\HG}[1]{\todo[textcolor=MidnightBlue,linecolor=Aquamarine,backgroundcolor=Aquamarine!25,bordercolor=Aquamarine!30,size=\tiny]{HG:~#1}}

\begin{document}
\title{Hybrid Planning with Receding Horizon:\\ A Case for Meta-self-awareness}
\author{\IEEEauthorblockN{
		Sona Ghahremani and
	Holger Giese}
	\IEEEauthorblockA{
		Hasso Plattner Institute, University of Potsdam, Potsdam, Germany, Email: \{firstname.lastname\}@hpi.de}
}

\maketitle

\begin{abstract}

The trade-off between the quality and timeliness of adaptation is  a multi-faceted challenge in engineering self-adaptive systems. Obtaining adaptation plans that fulfill system objectives with high utility and in a timely manner is the holy grail, however, as recent research revealed, it is not trivial. \emph{Hybrid planning} is concerned with resolving the time and quality trade-off via dynamically combining multiple planners that individually aim to perform either timely or with high quality. The choice of the most fitting planner is steered based on assessments of runtime information. A hybrid planner for a self-adaptive system requires (i) a decision-making mechanism that utilizes (ii) system-level as well as (iii) feedback control-level information at runtime.

In this paper, we present \app, a hybrid planner for self-adaptive systems. Inspired by model predictive control, \app leverages receding horizon control to utilize runtime information during its decision-making. Moreover, we propose to engineer \app for self-adaptive systems via two alternative designs that conform to \emph{meta-self-aware} architectures. Meta-self-awareness allows for obtaining knowledge and reasoning about own awareness via adding a higher-level reasoning entity. \app aims to address the problem of hybrid planning by considering it as a case for meta-self-awareness. %The benefits and limitations of the proposed designs for hybrid planners are investigated for two different case studies and in a comparison with a deterministic hybrid planner as the baseline. 

\end{abstract}

	\vspace*{-2mm}
\section{Introduction}
	\vspace*{-2mm}
	\noindent
Rapidly changing requirements, highly dynamic environments, and unpredictable operating conditions demand for runtime adaptation of software systems. Providing timely and high quality adaptation plans is the ultimate goal  of an adaptation manager. However, constructing a single automated adaptation policy that satisfies both of these conflicting requirements is challenging~\cite{filieri2011self}. \emph{Proactive} optimization-based policies often require an exhaustive search in the possible adaptation space which renders attaining optimal adaptation plans time-intensive~\cite{ Russell_Norvig2009}. Additionally, while \emph{reactive} condition-based solutions for adaptation deliver adaptation plans timely, they often fail to find the optimal solutions~\cite{de2017software}.

 \emph{Hybrid planning} employs control mechanisms where multiple adaptation policies are orchestrated to jointly carry out timely and optimal adaptations~\cite{pandey2017towards, diniz2020hybrid}. A hybrid planner for a self-adaptive system~(SAS) implements the ability to reason about different available adaptation policies, varying in the quality and timeliness of their plans, based on the runtime conditions. Time-critical operation conditions that demand timely rather than optimal adaptations benefit from employing cost-effective adaptation policies with short planning time. Under less time-sensitive operation conditions, optimization-based policies may provide high quality adaptation plans.  

While Control Theory has established mathematically grounded and practical frameworks for managing complex systems~\cite{MPCCoBra,SEmeetsCT}, it restricts the scope of the controllers to calculating set-points and prescribing required changes in the system input parameters~\cite{filieri2011self}. The black-box-oriented scheme of the Control Theory % where the controller only deals with system inputs and outputs 
further extends towards \emph{adaptive control}, where the controller may change its own control regime~\cite{landau2011adaptive}. This requires the controllers to have adjustable parameters. %Moreover, an \emph{adjustment mechanism} needs to be in place to oversee the parameter tuning via  layered arrangements of control loops, where the lower-level entities are controlled by the immediate higher-level loop~\cite{de2017software}.
 %However, the expectation from an \emph{adaptation} in the context of software engineering is distinguished from the one in the control engineering context~\cite{SEmeetsCT}. 
 %While the former considers adaptation as system's reaction to changing environmental and contextual conditions, the latter restricts the scope of the controller to calculating \emph{set-points} and prescribing required changes in the system input parameters accordingly~\cite{filieri2011self}. 
 %
 %\Note{Hierarchical and adaptive control}
 %the black box oriented scheme of the conventional control theory, where the controller only deals with system inputs and outputs can be further extended towards \emph{adaptive control}, where the controller may change its on control regime~\cite{landau2011adaptive}. This requires the controllers to have adjustable parameters. Moreover, an \emph{adjustment mechanism} needs to be in place to overlook the parameter tuning via  \emph{Hierarchical control}. In a hierarchical control scheme, i.e.~the layered arrangements of control loops, the lower level entities are controlled by the immediate higher level loop~\cite{de2017software}.
 \Note{Meta self-awareness}
In the realm of self-adaptive systems, adaptive control is perceived as reasoning about the adaptation logic~\cite{patikirikorala2012systematic}. The reasoning requires observing the behavior of the control loop in terms of effectiveness and performance, realizing the need for change, and prescribing the necessary decisions to steer the controller towards the desired behavior. %While techniques such as hierarchical control serve as a base to enable additional degrees of freedom in the adopted control loop~\cite{KramerandMagee2007}, by principle, they are limited in the extent of observation and modification they may demonstrate on the lower-level entities. 
\emph{Meta-self-awareness}, a notion surfacing only recently, captures the requirements for equipping self-aware systems with advanced self-reflective properties~\cite{Chen+Self-AwareSelf-ExpressiveSystems}. As a result, systems with meta-awareness properties can reason about changing trade-offs during their lifetime~\cite{SelfAware-Framework}. The control design and architecture of a meta-self-aware system builds on Control Theory as a prominent base, however, it extends the involvement \emph{scope} of the higher-level control loops in the lower-level entities.  

 There exists various research efforts in combining multiple adaptation solutions to fulfill the contradicting trade-offs in SAS~\cite{hrabia2019increasing,  trollmann2018hybrid}. \cite{diniz2020hybrid} proposes a hybrid approach that combines control theory principles with AI techniques to optimize the adaptation process.
\cite{sharifloo2016learning} presents a hierarchical hybrid planner where a learning-based policy adapts a rule-based policy to improve its decisions. % The individual policies have separate planning phases and learning based policy  optimizes the rules. hierarchical, if the adaptive system of one is part of the plan phase or adaptation problem of the other.
The hybrid planning approach in~\cite{pandey2016hybrid} operates an optimization-based planner in the background  while a deterministic policy adapts the system. Before each adaptation, the hybrid planner checks if the optimization-based policy can provide a plan.
\cite{vogel2010adaptation} presents a concurrent approach for self-adaptation in which a model of SAS is synchronized with multiple views on that model  and each partial model has its own adaptation mechanism.
% (e.g., component failure or performance).
Condition-based rules are used in~\cite{IQBAL2011871, bauer2018chameleon} to construct reactive hybrid planners. 
The main focus in providing hybrid solution for SAS so far has been set on developing individual adaptation policies such that, in coordination together, they cover a large spectrum of the solution space for SAS. In this paper, we focus on the orchestrating entity\ie~the realization of the decision-making mechanism within the hybrid planner. Our proposed solution has the characteristics of a generic hybrid planner since it considers the employed adaptation policies as a black-box, thus,  can coordinate arbitrary adaptation policies. 
 
We address the problem of hybrid planning by considering it as a case for meta-self-awareness. We present \app, a hybrid planner for self-adaptive systems. \app implements the planning phase as a controller conforming to the scheme of \emph{model predictive control}, thus,  leverages \emph{receding horizon} to utilize runtime information and adjust its control parameters at runtime. To engineer \app for SAS, we propose two alternative designs, \emph{external} and \emph{internal}, that conform to \emph{meta-self-aware} architectures. The designs build on the framework for realizing meta-self-awareness in the architecture from our earlier work~\cite{SelfAware-Arch1}.

 We study the effectiveness of the designs for hybrid planning by answering the following Research Questions (RQ). \reqone~how do internal and external designs for meta-self-awareness affect \app? \reqtwo~how does \app perform in comparison to a deterministic hybrid planner? \reqthree~what are the effects of  hybrid planning on the quality and timeliness of the adaptation? We show that meta-awareness capabilities, realized either by the external or internal design, are beneficial for hybrid planning as they provide extended control flexibility at runtime. %Moreover, our experiments  suggest that the common practice of increasing the visibility of control loops in the architecture is also beneficial in designing meta-self-aware systems with hybrid planning. 

 %Our scheme adds an additional control entity on top of the SAS. The additional level, can be realize at the same level as the original MAPE loop. Alternatively, the meta control  loop can be considered external to the SAS. We evaluate the effect of adding meta self-awareness capabilities via comparing them to a baseline scenario where such capabilities are discarded. Our preliminary experiments on a SAS suggest that adding meta self-awareness to a SAS leads to improvements of the adaptation performance. The results show that realizing meta self-adaptation via an external loop is specifically beneficial as it operates on larger time scale compared to the internal alternative. The external meta loop makes more informed decisions based on history that is, its accumulated observation and experience over longer periods. Consequently, the meta loop makes more resilient decisions avoiding nervous behavior.
 
 \Note{tour} 
Section~\ref{sec:preReqs} discusses the prerequisites. A motivating example  is presented in~Section~\ref{sec:CaseStudy}. Section~\ref{sec:hypzon} presents \app and~Section~\ref{sec:MetaSelfAware} presents the design and application of \app as a case for meta-self-awareness. The \textbf{RQ}s are investigated in~Section~\ref{sec:Evaluation} and Section~\ref{sec:conclusion} concludes the paper.% and provides future work.%discusses adaptation concerns demanding different reasoning levels and  Section~\ref{sec:MetaSelfAware} 

%\cite{de2017software} on Adaptive and Hierarchical Control.

%\newpage

\section{Prerequisites}\label{sec:preReqs}
%In this section, we first present a basic rule-based adaptive system, followed by a description of \emph{static} vs \emph{dynamic} adaptation policies. The section also presents concepts for meta-awareness enabled in the introduced systems. Finally here we cover some basic background on model predictive control in the context of self-adaptive systems.

\subsection{Self-adaptive Systems}\label{subsec:RBSAS}
% 
%The architectural attributes of the system constitute the \emph{configuration space} of SAS. A configuration of SAS at a given point in time is identified as a vector of configuration settings $\bar{c}$.
%
%The \emph{observable attributes} are monitored at runtime and reflect the impact of the environment on the system or its context. 
%
%An observation is a vector of observation values $\bar{o}$. The current observations $\bar{o}$ may change to a new observation $\bar{o}'$ independently of the configuration changes.
%
%A pair of a configuration and the corresponding observations $(\bar{c},\bar{o})$ constitute the \emph{state} of a SAS at each point in time~\cite{brun2013design}.
%
%Both $\bar{c}$ and $\bar{o}$ together denote the design space of a SAS~\cite{Blair+2009}.
The execution of an \emph{adaptation action} $a\in A$~with~$A$ the set of available actions, results in adapting the SAS from state $s$ to $s'$ with~$s$ and $s' \in S$ and $S$ the \emph{state space} of the SAS. 
%
%and are guarded with \emph{conditions}. After the adaptation engine detects a change \emph{event}, if any of the adaptation rule conditions are satisfied, the rules become applicable as an adaptation action to adapt the system accordingly and resolve the \emph{adaptation issue}. Adaptation rules conform to the general class of \emph{event-condition-action} rules. 
%
%\textit{Static and Dynamic Adaptation Policies.}\SG{rename}
%Although numerous definitions of  \emph{policy} have been put forward in recent years, the most common notion of 
A \emph{policy} is generally defined as a set of control decisions that map states to actions~\cite{Russell_Norvig2009}. An \emph{adaptation policy}~$\pi$ represents an encapsulation of the system's adaptive behavior governing the choice of adaptation actions when applicable. For each state~$s$, $\pi(s)$ indicates the adaptation action~$a$ to be executed, i.e.,~$\pi(s)=a$. 
%The building blocks of adaptation policies are \emph{perceptions} and \emph{responses}. The perceptions rely on system's ability for awareness. Responses are implemented as adaptation actions in accordance with (or triggered by) the perceptions.   
%
%For a pair $(s,a)$, where the effect of the actions are deterministic, the execution of action $a$ in state $s$ transitions the system to a next state $s'$. 
%
%evaluates an objective function $U$, known as \emph{utility function}. 
 $EU(s)$ represents the expected utility of $s$ i.e., a scalar value as a quality metric that identifies the degree to which system goals and requirements are satisfied in $s$. %assuming that the system transitions from configuration $c$ to $c'$ if action $a$ is executed.
%
%In presence of nondeterministic operation conditions, for any possible next state $s'$, it only holds that $0<P(s'|s,a)\leq1$.
%This is similar to the \emph{state-action value function} in multi-agent systems context~\cite{}. 
For all the applicable adaptation actions $a\in A$ in state $s$, an \emph{optimal} $\pi(s)$ chooses the action that maximizes the expected utility of the subsequent state $s'$\ie~$EU(s')$.

The utility of a system can be defined as a function of the system \emph{quality} \emph{dimensions}. A common practice to obtain a representative utility function is via the system's Service Level Agreements (SLA's) where the business preferences define the system \emph{objectives}. A multi-objective utility function is an aggregation of multiple quality dimensions where each dimension represents  a business objective.\emph{ Utility sub-functions} are employed to assign values to the dimensions~\cite{WalshAutonomicComputing}.

A \emph{static} adaptation policy maps an action $a$ to state $s$ based on design-time estimates for the resulting state~$s'$. Therefore, the estimations for $EU(s')$ are agnostic to using any runtime observation. %Rainbow~\cite{2012ChengStitch} is an example of a static adaptation policy as it only uses configuration attributes $\bar{c}$ to \emph{estimate} the expected utility for the action.
%
%refers to a class of adaptation policies where for all \emph{possible} combinations of ($s$, $a$), the resulting $s'$, $EU(s')$, and consequently $\pi(s)$ are determined statically, that is, the estimations for the expected utility agnostic to using any runtime observation $\bar{o}$. Rainbow~\cite{2012ChengStitch} is an example of a static adaptation policy as it only uses architectural properties of the system for utility estimation statically. 
%
In contrast, a \emph{dynamic} adaptation policy leverages runtime observations, available during system execution, to \emph{compute} expected utility values for the applicable actions.%~\cite{SG_ICAC}.

\subsection{Hybrid Planning for SAS }\label{subsec:HybridPlanning}
Hybrid planning for SAS in general refers to combing two or more adaptation policies to adapt the system. More specifically, a \emph{Coordinating Hybrid Planner}~(CHP) combines adaptation policies that (i) target the same SAS and (ii) the same \emph{planning problem} but (iii) keeps their planning phases separate. The planning phase of a SAS employing a CHP can be subdivided into the original planning phases from the combined policies~\cite{trollmann2018hybrid}.
An \emph{adaptation issue} refers to a deviation from the desired state and indicates that the system, in its current state $s$, requires an adaptation. A \emph{planning problem} is a set of adaptation issues that simultaneously affect the SAS. For a {planning problem}, a \emph{plan} is an ordered list of adaptation actions that each action resolves at least one of the adaptation issues in the planning problem. The \emph{look-ahead horizon} is the time steps into the future that are considered during planning. A \emph{planning horizon} $\Phi$ is a prefix of the look-ahead horizon that is planned for. A planning horizon of size $|\Phi|\le l$ with $l$ the size of the look-ahead horizon, only plans for $|\Phi|$ out of $l$ adaptation issues in the look-ahead horizon. An infinite planning horizon allows for considering the entire look-ahead horizon for planning. The \emph{utility of plan} is the expected utility of the target system state after the execution of the adaptation actions constituting the \emph{plan}. 
\begin{figure}[t]
	\vspace*{-6mm}
	\begin{centering}
		\includegraphics[width=.99\linewidth]{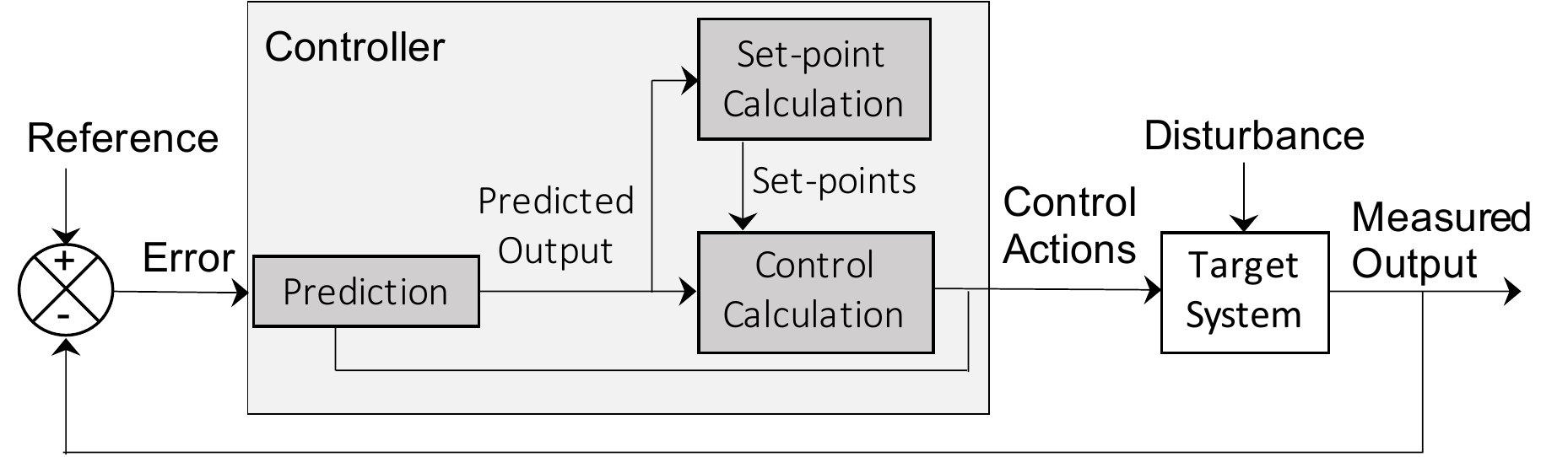}
		\vspace*{-4mm}
		\caption{Block diagram of a feedback control system. The gray box extends the controller to illustrate a model predictive controller. }
		\label{fig:controlloop}
	\end{centering}
	
	\vspace*{-3mm}
\end{figure}
 
A CHP is concerned with two aspects of the available policies: quality and timeliness. For a given planning problem, the quality of a policy~$\pi_i$ is quantified as the expected utility of the plan provided by~$\pi_i$\ie~$EU(plan_i)$. For timeliness, an estimation of the time required by~$\pi_i$ to provide a plan\ie~$ET(plan_i)$, is required~\cite{pandey2017towards}. Such estimations can be obtained via  theoretical modeling such as employing worst-case time models~\cite{pandey2016hybrid} or based on empirical profiling.
	\vspace*{-1mm}
\subsection{Feedback Control for SAS }\label{subsec:control}
A control feedback loop that governs a SAS observes the system output in \emph{sampling intervals}~$\mathcal{I}$ and adapts the system towards its \emph{set-points} to prevent violation of the system requirements and goals over a \emph{prediction horizon}. 
 The \emph{target system}\ie~SAS is controlled via the \emph{control actions} from the\emph{ controller}--see~Fig.~\ref{fig:controlloop}.
The \emph{reference} input is the desired value of the system’s \emph{measured output}. The goal is that, despite the \emph{disturbance} affecting the target system, the measured output is sufficiently close to the reference. For this purpose, the difference
between the measurement and the reference, i.e., the \emph{error} is fed back to the controller to determine the control actions required to achieve the reference value.

\emph{Model Predictive Control}~(MPC) is a technique that formulates a multi-variable optimization function,~e.g.,~a utility function, to generate \emph{set-point}s. Set points define the target values for \emph{control calculations}.  Control calculations determine a sequence of $M$ control actions,~i.e.,~\emph{control horizon}, such that the \emph{predicted output} moves towards the set-points over a finite \emph{prediction horizon} $P$~\cite{ghanbari2014replica}. %The control actions are the required modifications to the input variables and account for all possible trajectories up to a the calculated \emph{ control horizon} $M$~\cite{kusic2009power}.
The number of the predictions $P$ is referred to as the \emph{prediction horizon} while the number of the control actions $M$ is called \emph{control horizon}. 
The MPC \emph{receding horizon control} suggests that although a sequence of $M$ control actions is calculated at each time point, only the \emph{first} action is executed. A new sequence is calculated after new observations become available.
%A distinguishing feature of MPC is its receding horizon approach. Although a sequence of $M$ control actions is calculated at each time point, only the \emph{first} action is executed. Then a new sequence is calculated at the next sampling instant, after new observations become available.%; again only the first action is implemented. This procedure is repeated at each sampling instant. %An $M$-step control strategy calculated and only the first step is implemented.
%
Employing a receding horizon of size \emph{one} supports the case where  the variables available for the control calculations change from one execution time to the next. If the control structure changes from one control execution time to another, but the MPC controller does not recalculate the parameters, the subsequent control calculations may become \emph{ill-conditioned}~\cite{seborg2010process}. 
 For a SAS, before each adaptation, the adaptation policy $\pi$ optimizes an objective function to select the adaptation actions that maximize the said function. MPC has been proven effective in formulating the optimization problem where the control actions represent adaptation actions and the targets are the system states with maximum $EU$~\cite{angelopoulos2018engineering}. 
%\vspace*{-1mm}
\subsection{Self-awareness and Meta-self-awareness }\label{subsec:selfawareness+Meta}
%\vspace*{-1mm}
\Note{self-awareness}

A \emph{self-aware} system is identified by two main characteristics. First, the ability to \emph{learn} models capturing knowledge about the system, its context, and its goals on an ongoing basis. Second, \emph{reasoning} based on the models for analysis and planning concerns. Computational self-awareness is achieved via a Model-based Learning, Reasoning, and Acting loop (LRA-M loop). There can be multiple variations to Acting, e.g.,~explaining, reporting, suggesting, and adapting. Self-adaptation, realized via a MAPE-K feedback loop, is one of the advanced characteristic of the self-aware systems where the \emph{scope} of \emph{Acting} is set to \emph{Adapting}~\cite{SelfAware-Framework}.  As a result, a system realizing MAPE-K loop becomes aware of itself and its context. 
The \emph{object} of the awareness is the entity being reasoned upon. The \emph{subject} of the awareness is the entity performing the reasoning. In the following, we target self-aware systems with adaptation capabilities. % that conform to an LRA-M architecture (see lower-level loop in~Fig.~\ref{fig:SelfAw}).
%

%Similar to the MAPE-K feedback loops for SAS, the LRA-M loop is addressed during the architectural design of self-aware computing systems. This allows engineers to explicitly decide and reason about the system’s self-awareness capabilities.

\Note{meta-self-awareness}
A \emph{meta-self-aware} system can obtain knowledge about its own awareness and how it is exercised. %and the cost and benefit of maintaining different awareness levels
 A higher-level self-aware entity\eg~a MAPE-K control loop, \emph{reflects} on the benefits and costs of maintaining increased awareness as well as the capacities for it~\cite{Chen+Self-AwareSelf-ExpressiveSystems}.
Meta-self-awareness is concerned with two classes of objects; The elements of the lower-level awareness loop,~e.g.,~a learning process or an adaptation logic, and the output of these elements, i.e.,~the models or specifications produced by the object being reasoned upon~\cite{SelfAware-Framework}. In order to explicitly capture the meta-self-awareness properties in the architectural design, meta-self-awareness can be realized either as a built-in capability or as an external meta-awareness layer~\cite{SelfAware-Arch1}--similar to the internal and external approaches for engineering systems with self-adaptation properties~\cite{cheng2004making}.

%\vspace*{-2mm}
 %A meta self-aware system can adapt the way in which it realizes an awareness level, for example, by changing algorithms for realizing that level or by deciding whether or not to employ that level at all~\cite{Lewis+15}. 
%
%

%
%\newpage
\section{Motivating Example}\label{sec:CaseStudy}

\begin{figure}[t]
\vspace*{-6mm}
\begin{centering}
	\includegraphics[width=.6\linewidth]{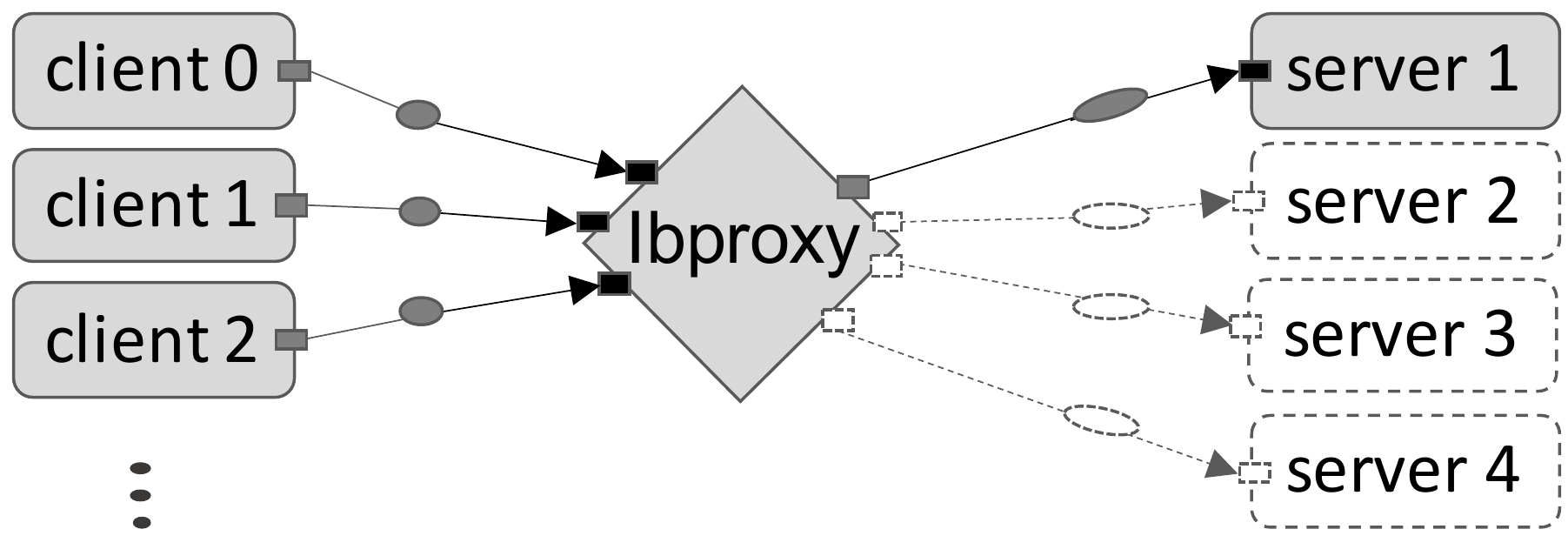}
	\vspace*{-2mm}
	\caption{\z~ architecture. Dashed lines represent available but inactive elements.}
	\label{fig:znnArch}
\end{centering}

\vspace*{-4mm}
\end{figure}

\noindent
As a running example we employ \z~~\cite{2012ChengStitch} that simulates a news service website. Clients make content requests to one of the servers. A load balancer distributes requests across a server pool. The size of the pool can be dynamically adjusted to balance the server utilization against the service response time.
\z~ is a web-based client-server system conforming to an N-tier style--see~Fig.~\ref{fig:znnArch}. %, a load balancer distributes requests across a server pool with a dynamically adjustable size to balance server utilization against service response time.
 Certain system and client information such as  server load, request response time, and the connection bandwidth can be monitored. The goal of \z~~is to provide short response times to the clients while keeping the cost of the server pool within the budget. Four types of adaptation issues may affect \z~: \late~~in the response times, \budget~ cost of adding servers to the pool, \utiliz~ servers, and \quality~ contents. There are four adaptation actions applicable to modify the configuration of \z~: \discharge,  \enlist~, \increase~, and \reduce~ of the content.
Response time, content quality, server utilization, and budget are the four objectives of \z~. These objectives are captured by four quality dimensions respectively. Each dimension is represented by a utility sub-function as described in~Table~\ref{tab:znnUtility}--see Section~\ref{subsec:RBSAS}. %that can either be observed~(e.g.~response time) or obtained from the system configuration~(e.g.~server load). Utility values are assigned to particular values of quality dimensions.
 %
 % Cheng~et~al.~\cite{2012ChengStitch} employ a discrete custom utility function in the form of $(x_i, y_i)$ where the utility $y_i$ is assigned to the value $x_i$ of the quality attribute. For example the discrete utility function $(low, 1)$ for the quality attribute Response time means that low response time obtains a utility value equal to $1$. 
% We Extend the discrete custom utility functions introduced by Cheng~et~al in~\cite{2012ChengStitch} for \z~~and replace them by linear utility sub-functions as described in~Table~\ref{tab:znnUtility}.% describes the quality dimensions and the employed utility sub-functions. %Different quality dimensions have different impacts on the overall utility of the system.  Each quality dimension $i$ affects the overall utility with a weight of $w_i$ such that $ \sum_{i}^{} w_i = 1 $.%
 
%
%Similar to~\cite{2012ChengStitch}, we use predefined weights as described in Table~\ref{tab:znnUtility} for the static adaptation policy. However, dynamic policy computes the weights before each utility computation based on the runtime operation conditions. 
%
%RT
$RT$ in Table~\ref{tab:znnUtility} is the estimated client response time and $RT_{max}$ is set to $90$ seconds, that is when \z~ throws a \emph{request timeout} exception and ends the session.
  \begin{table}[b]\centering
	\vspace*{-5mm}
	\caption { Utility Sub-functions for \z~~}
	\label{tab:znnUtility} 
	\vspace*{-2mm}
	\resizebox{7cm}{!}{
		\begin{tabular}{lllr}\toprule
			\textbf{ID}&\textbf{Quality Dimension}&\textbf{Utility Sub-function}&\textbf{$w_i$}\\
			\cmidrule{1-4}% \cmidrule{3-4}	
			
			$u_R$ &Response time&$u_R=1-\frac{RT}{RT_{max}}$&0.4\\
			$u_Q$& Content quality&$u_Q=~Server.quality$&0.2\\
			
			$u_U$	&Server utilization&$u_U=~Server.utilization$&0.1\\
			
			$u_C$& Cost&	$u_C=Server.cost$&0.3\\
			\bottomrule
		\end{tabular}
	}
	\vspace*{-6mm}
\end{table}
A server in \z~ can transfer content with three different qualities that are quantified as: (low,~$0$), (medium,~$0.5$), and (high,~$1$). Therefore,  $Server.quality$  can have one of the $0$,~$0.5$, or $1$ values. $Server.utilization$ is the percentage of the server capacity that is in-use, and finally,  $Server.cost$ is the operational cost of the server which can vary for different providers. 
\begin{equation}\label{eq:Znnutility}
	U_{znn}(s)=\sum_{client}^{}w_ru_R\, +\sum_{server}(w_q^{}u_Q\, + w_uu_U\,- w_cu_C)
\end{equation}

Similarly to~\cite{2012ChengStitch}, we define the overall utility of the system as a weighted sum of the utility sub-functions. The weights $w_i$ in Table~\ref{tab:znnUtility} are extracted from~\cite{2012ChengStitch}.  For each state $s$, the overall utility of \z~~is defined according to~(\ref{eq:Znnutility}). For each target state $s'$, the expected utility of the state is an estimation of~(\ref{eq:Znnutility})\ie~$ EU(s')=	\hat{U}_{znn}(s')$.

%We equipped \z~~with self-adaptive properties via adding an LRA-M loop, henceforth \emph{adaptation loop}, to the system. The adaptation loop employs an adaptation policy $\pi$ that exhibits the functionality of a SAS, that is, resolving adaptation issues at runtime. To this end, $\pi$ selects the adaptation action $a\in\,A$ that is expected to resolve the adaptation issue while maximizing $EU$. \late~, \utiliz~,~\quality~, and \budget~~constitute the potential adaptation issues in \z~. \reduce~, \enlist~, \discharge,~and \increase~~embody the primitive actions in~$A$. 

\section{Hybrid Planning with Receding  Horizon}\label{sec:hypzon}
\noindent
In this section, we present \app, a coordinating \textsc{Hy}brid \textsc{p}lann\textsc{e}r for SAS employing receding hori\textsc{Zon} control. \app aims to address the planning problem in a SAS at runtime by considering multiple adaptation policies and selecting based on the time and quality objectives. \app implements the planning phase of a MAPE-K  loop.
During an adaptation cycle, once the analysis phase has detected the need to plan an adaptation\ie~there exists a planning problem--see Section~\ref{subsec:HybridPlanning}--\app takes the system state, the set of available adaptation polices, and estimations of $EU(plan)$ and $ET(plan)$ for the available policies. Then, the planner decides which policy best suits the current operation condition.
\begin{figure}[t]
	\vspace*{-6mm}
	\begin{centering}
		\includegraphics[width=.8\linewidth]{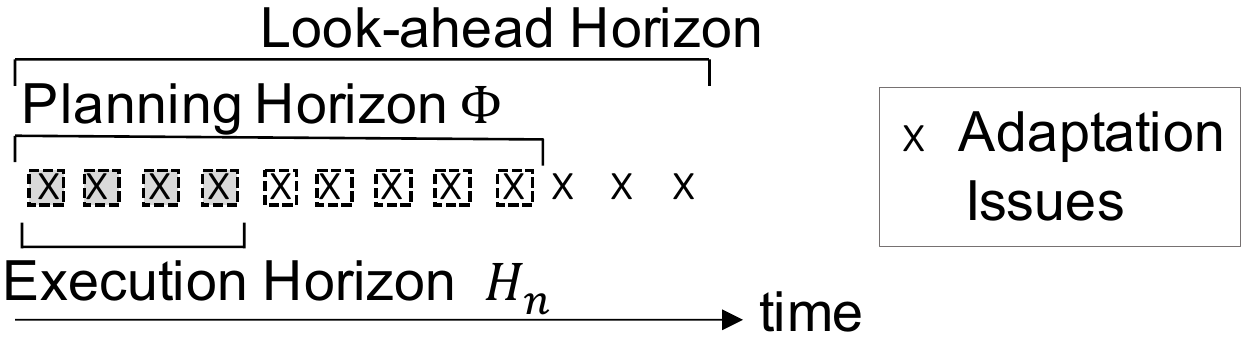}
		\vspace*{-2mm}
		\caption{Look-ahead, planning and execution horizon in \app. }
		\label{fig:Horizon}
	\end{centering}
	
	\vspace*{-4mm}
\end{figure}

\app implements the planning phase as a controller conforming to the scheme of  model predictive control with receding horizon. The MPC-based planner, at each sampling instance~$\mathcal{I}$, makes new measurements, and based on the operation condition, decides if a policy switch is required.
The process of resolving a set of adaptation issues via selecting a set of adaptation actions that maximize a utility function is formulated  as an MPC receding horizon control problem.  %The look-ahead horizon in hybrid planning is captured by the prediction horizon in MPC. Moreover, the planning horizon of the hybrid planner can be implemented via the control horizon of a model predictive controller. 
\app implements a hybrid planner as an MPC controller via mapping the look-ahead and planning horizons of the planner to the prediction and control horizons of an MPC controller respectively--see~Section~\ref{subsec:control}.

 \app extends the notion of MPC receding horizon of size \emph{one} to an \emph{execution horizon} $H_n$ with adjustable size $n$. Fig.~\ref{fig:Horizon} shows an example of a look-ahead horizon, planning horizon, and execution horizon in \app--see~Section~\ref{subsec:HybridPlanning}.
%
%The size of the execution horizon, $n$, is a parameter that in addition to MPC, is relevant for sequence planning \cite{wongpiromsarn2012receding}, and planning phase of various approaches (e.g.,~\cite{li2015intermodal}) that employ  scheduling strategies for the potential adaptation steps. 
As explained in Section~\ref{subsec:control}, in MPC, the control horizon is a list of \emph{all} the actions that are planned during an adaptation cycle. This is captured as the planning horizon~$\Phi$ in \app.  % that have been assigned to the detected adaptation issues.  %assigned to all the observed failures before the current self-healing loop. 
%During the planning phase of a self-healing loop, self-healing approaches assign the proper repair actions to the detected failures. The assignment and ordering of the repair actions is steered based on design-time or runtime estimates of the impact of the repair actions (cf.~Section~\ref{subsec:approaches}).
%
%a  considers switching the adaptation policies to solve the remaining issues based on the current operation condition.
%
%For example, if the execution horizon is of size one~(i.e.,~$H_1$), only the first planned action is executed and the re-planning is initiated in the subsequent adaptation cycle based on the new observations and remaining unresolved issues. 
%
%The execution horizon is a subset of length $n$ of the control horizon.	
%	\pi(s)=\argmax_a EU(s'| P(s'|s,a)>0)
As stated by Pandey \et~\cite{pandey2017towards}, it is difficult to verify the compatibility between the plans of different policies,  how to choose the planning horizon, and when to stop using one plan and switch to another policy for planning. In the following, we describe how \app addresses these challenges.
%
 %The \emph{adjust} method in Algorithm~\ref{Algorithm:hypzn} 
 \app uses runtime information such as the planning time of the adaptation policies,  system load, number and type of the adaptation issues, and the cost of switching between the policies to decide on the size of the planning and execution horizons. Employing planning and execution horizons with adjustable size provides for runtime flexibility. Before choosing a policy, \app adjusts the size of the planning horizon~$\Phi$ with respect to its estimation of the policy planning time\ie$ET(plan)$. For example, if \z~ is affected by large numbers of \late~~issues as well as \quality~~issues, \app may restrict the planning horizon to first consider the more critical issues\ie~\late~. This way, \app reduces the expected planning time by reducing the size of~$\Phi$, thus, the \late~~issues are resolved relatively faster.

An execution horizon $H_n$ only considers the first $n$ adaptation actions in the planning horizon for execution in the current adaptation cycle. After executing the $n$ adaptation actions, \app stops the execution of the plan and the remaining unresolved issues are considered together with the newly observed issues in a subsequent adaptation cycle as a new planning problem. %At the end of each execution horizon, \app makes new observations\ie~sampling, and checks if a policy switch is required.
Employing execution horizon of small size in \app~results in utilizing the most recent adaptation issues immediately. In contrast, large execution horizons ignore the recent observations until all the actions in the planning horizon are executed--see Fig.~\ref{fig:Horizon}. Small sizes for $H_n$ demand more frequent planning. Moreover, the execution of the actions that are in the planning horizon and not in the execution horizon is postponed to the subsequent adaptation cycle(s). In cases where the planning phase of the adaptation policy has a large overhead, frequent planning might affect the adaptation time negatively. 
%After executing the $n$ actions, \app reconsiders policy switch. 
 When \app switches the adaptation policy, the employed policy plans for the remaining adaptation issues whose corresponding adaptation actions in the planning horizon were not included in the execution horizon. Moreover, the  employed policy also considers the newly detected adaptation issues. Consequently, after each policy switch, the planning problem is considered anew during the control calculation in \app. This way, \app guarantees that after a policy switch, the active policy calculates the plan according to the most recently observed conditions while taking into account the already existing issues.
 
 In order to guarantee the compatibility between a plan and the planning problem, \app only executes one policy at a time and avoids concurrent executions of multiple policies. As a result, a planning problem that is assigned to a policy remains unchanged during the planning time. This way, once the plan is ready, \app does not check if the plan is still applicable to the current planning problem. This feature in \app avoids the runtime overhead that is caused by compatibility analysis between the planners. However, concurrent executions of planners may reduce the time that the hybrid planner has to wait until a plan is ready.% as it allows for executing proactive policies in the background. 
% of time. 
%	\vspace*{-2mm}
\section{\app: Design and Application}\label{sec:MetaSelfAware}
\noindent
%In this section we propose two alternative designs to engineer~SAS~with meta-self-awareness properties and discuss their application to resolve a selection of \emph{runtime concerns} in the context of self-adaptation.% We elaborate on the benefits and the limitations of each design. 
In this section, we argue that equipping a SAS with hybrid planning should be realized as a meta-self-awareness property. To this end, building on our previous study towards making meta-self-awareness visible in the architecture~\cite{SelfAware-Arch1}, we propose two designs to engineer a~SAS~with meta-self-awareness properties and  show how the two designs are realized in \app. The self-awareness capabilities are realized via MAPE-K loop--see Section~\ref{subsec:selfawareness+Meta}.

%preliminary categorization of \emph{four} potential \emph{Adaptation Concerns}~(AC) that a SAS encounters during its execution and a discussion on how each of the design variants equip the basis SAS to address the concern.

%, underlining the need for a disciplined engineering practice in this area.
% \SG{We build on the architectural study of Giese et~al.in\cite{SelfAware-Arch1} }
%
%\vspace*{-2mm}
\subsection{Meta-self-aware Designs Realizing Hybrid Planning}\label{subsec:design} 
\begin{figure}[t]
	\vspace*{-6mm}
	\begin{centering}
		\includegraphics[width=0.85\linewidth]{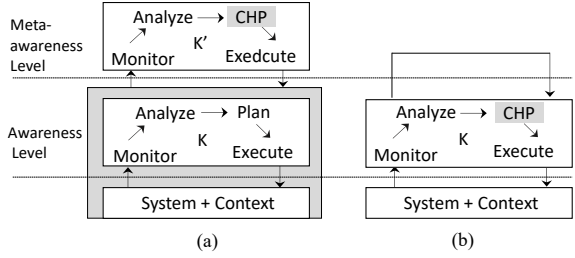}
		\caption{External~(a) and internal~(b) design for meta-self-awareness.}
		\vspace*{-2mm}
		\label{fig:MetaSelfAw}
	\end{centering}
	\vspace*{-4mm}
\end{figure}

\noindent\textbf{External design}
In order to explicitly separate the awareness and meta-awareness levels, as depicted in~Fig.~\ref{fig:MetaSelfAw}-(a), two MAPE-K loops are employed. 
The  loop at the \emph{meta-awareness level} implements the CHP. The higher-level loop observes the awareness level in combination with the system and context, reasons about them, and adapts them accordingly. 

%Models are awareness models that are evolving at runtime. Act is translated to adapt`
%

 \Note{General discussion on characteristics}
  
  %The External design supports explicit separation of concerns (i.e.,~awareness and meta-awareness) at the architecture level and allows for reusability, easier maintenance, and  independent evolution of each level. As a result, separate and independent mechanisms for observing, learning, and reasoning logic may be employed by  each level.
  %
  %Realizing meta-awareness via an external level provides for a global view on the meta-awareness object. This allows for observing phenomena with global scope that are not observable at the awareness level. %via a self-loop.
  %z
  The external design operates the CHP on a different timescale than the lower-level MAPE-K loop. The lower-level loop is executed more frequently to guarantee timely adaptation concerning the part of the system under its direct control. The meta-awareness loop however, operates at a relatively larger timescale since it is inherently concerned with relatively more sparse phenomena to react upon~\cite{de2013software}.
   %This is both due to the sparsity of the relevant concerns (compared to adaptation issues that are more frequent than other concerns) and its more coarse grained invocation intervals by design.
%

\noindent\textbf{Internal design} An internal realization of the meta-self-awareness properties is possible via employing an  \emph{awareness self-loop}--see~Fig.~\ref{fig:MetaSelfAw}-(b). In this design, the MAPE-K loop \emph{observes} and \emph{affects} itself. %~and serves both as the awareness subject and the meta-awareness subject.
 The subject and object of meta-awareness are not architecturally separated, consequently, one element,~i.e.,~the MAPE-K loop, performs both the \emph{reasoning} and \emph{being reasoned about} parts of the meta-awareness. In the internal design, the awareness level is also aware of itself.

% \Note{General discussion on characteristics}
The internal design implements the awareness and the meta-awareness properties in an intertwined manner and keeps the subject of the meta-awareness close to the object. In this design, the meta-awareness self-loop, hence the CHP,  operates at the same timescale as the awareness loop.% since it is executed at the same level. 

In the proposed designs, the focus is on the \emph{functional} aspects enabled by each design rather than the \emph{architectural} aspects. %\eg the architectural components realizing the elements of each designs. 
However, the architectural decision of separating the awareness and the meta-awareness loops in the external design and combining them in the internal design steers the decisions relevant to  the functional aspect\ie~the sampling and execution intervals of the CHP in each design.

%	\vspace*{-2mm}
\subsection{Application}\label{subsec:application}
\app is realized as a  meta-awareness subject via both external and internal designs. The variants are called \extr~and \intr~respectively. The external design, thus, \extr operates the  meta-awareness loop in coarser time intervals compared to the awareness loop. Consequently, compared to \intr, the MPC controller in \extr has larger sampling intervals ($\mathcal{I}$)--see~Section~\ref{subsec:HybridPlanning}. 
Algorithm~\ref{Algorithm:hypzn} shows a high-level description of \app. 

\begin{algorithm}[h]
	\small
	\caption{Hybrid Planning with \app}
	\label{Algorithm:hypzn}
	\begin{algorithmic}[1]
		\Require s, $\Pi$, $RT_{\mathcal{I}}$
		\State $\Phi\gets$ look-ahead horizon
		\If{ $RT_{\mathcal{I}}$ \textbf{optimal}} 
		\State $\pi_{curr}=	\pi_i\in\Pi \ni i=\displaystyle{\argmax_i EU(plan_i)}$
		\State $H_n\gets\infty$ 
		\ElsIf { $RT_{\mathcal{I}}$ \textbf{inRange}} 
		\ForAll{$\pi_j\in\Pi$}
		\If {$c_{curr,j} +ET(plan_j)+RT_{\mathcal{I}} $ \textbf{inRange} \textbf{AND}  $j=\displaystyle{\argmax_j EU(plan_j)}$} 
		\State $\pi_{curr}=\pi_j$
		\State \textbf{adjust}($H_n$) 
		\EndIf
		\EndFor
		\If{$\pi_{curr}=null$} 
		\State  \textbf{adjust}($\Phi$) 
		\State Go to  3
		\EndIf
		\ElsIf { $RT_{\mathcal{I}}$ \textbf{high}} 
		\State $|\Phi|\gets$ 1
		\State $H_n\gets$ 1
		\State $\pi_{curr}=\pi_j\in\Pi \ni j=\displaystyle{\argmin_j ET(plan_j)}$ 
		\EndIf
		\State List of Actions $ \gets \pi_{curr}(s)$
		
		\Return  List of Actions , $H_n$
		
	\end{algorithmic}
	%	\vspace{-0.4cm}%
\end{algorithm}%

\app is concerned with (i)~\emph{control parameter tuning}\ie~tuning the size of the planning and execution horizons, and~(ii)~\emph{policy switch}. %\ie~deciding to switch between the available adaptation policies. 
 As depicted in Algorithm~\ref{Algorithm:hypzn}, \app requires system state $s$ and the set of available adaptation polices $\Pi$. $\Pi$ also includes estimations of $EU(plan)$ and $ET(plan)$ for the available policies. The planning horizon~$\Phi$ is initially set to fully include the look-ahead horizon\ie all the the existing adaptation issues--see~Fig.~\ref{fig:Horizon}.
  In addition to the information available as system state $s$\eg current system load and utility, \app also maintains the average response time of the SAS during the sampling interval~$\mathcal{I}$\ie$RT_{\mathcal{I}}$. Based on the specific business objectives of SAS, \app defines three ranges for the average response time: \emph{optimal}\ie~$RT_{\mathcal{I}}$ is below a minimal threshold, \emph{inRange}\ie~$RT_{\mathcal{I}}$ is within the acceptable range, and \emph{high}\ie~$RT_{\mathcal{I}}$ is higher than the permitted upper bound. These thresholds are subject to change at runtime to reflect the changing goals and requirements. 
\newline\textbf{~Control parameter tuning}
 For a given planning problem, both \app variants employ the method \emph{adjust}--line~9 and 11 in~Algorithm~\ref{Algorithm:hypzn}--to tune their control parameters at runtime. The method uses estimations of the $EU(plan)$ and $ET(plan)$ for the policies, the current and estimated system load, and the number as well as the type of the adaptation issues.
\newline\textbf{~Policy switch}
Switching between policies has a cost as it requires deploying specific settings for the new policy,~e.g.,~initializing a constraint solver or loading prediction models. The switch from policy $\pi_{i}$ to $\pi_{j}$ is charged with a cost $c_{ij}$ that is subtracted from the system utility. \app reasons about the trade-off between cost and benefit of the switch at runtime--line 2, 5, and 13 in~Algorithm~\ref{Algorithm:hypzn}.
If $RT_{\mathcal{I}}$ is \emph{optimal}, \app switches to the policy with highest expected plan utility and executes the full plan--line~2-4~(index \emph{curr} represents the current choice). If $RT_{\mathcal{I}}$ is \emph{inRange}, \app searches for a policy with the highest $EU(plan)$ such that the sum of the policy switch cost, $ET(plan)$, and $RT_{\mathcal{I}}$ is still \emph{inRange}--line~5-8. \app uses $RT_{\mathcal{I}}$ as an estimate for $RT$ during the next interval.
In case \app does not find a match, the size of the planning horizon is reduced until \app finds a match--line~10-12. Finally, if $RT_{\mathcal{I}}$ is \emph{high}, \app sets the size of the planning and execution horizons to one and searches for a policy with the minimum planning time--line~13-16. 

In both variants, decisions for tuning the control parameters and policy switch is made based on the average values over a sampling interval~$\mathcal{I}$. Therefore, \extr makes estimates of the $EU(plan)$, $ET(plan)$, and system load based on the average values of observations over a longer monitoring period compared to \intr. Thus, \extr~collects accumulated observations of system executions that can be used to estimate the  operation condition during the next interval. \intr however, operates more frequently and decides based on the observations over a relatively smaller~$\mathcal{I}$.

%
%
%\newpage
\vspace*{-2mm}
\section{Evaluation}\label{sec:Evaluation}

\noindent
%This section presents an evaluation of the \extr~~and~\intr~~variants based on the RTCs as discussed in~Section~\ref{sec:application}.
In this section, we evaluate the application of the \app variants on~\z~. The experiments are designed to answer the three Research Questions. \reqone~how do internal and external designs for meta-self-awareness affect \app? \reqtwo~how does \app perform in comparison to a deterministic hybrid planner? \reqthree~what are the effects of  hybrid planning on the quality and timeliness of the adaptation? 
\vspace*{-1mm}
\subsection{Case Study and Deterministic Hybrid Planner}\label{subsec:caseStudyEval}
\noindent\textbf{Case study}~The employed case study is \z~~from Section~\ref{sec:CaseStudy}. The request arrival traces are generated based on the commonly used (e.g.,~\cite{7572219}) web traffic logs of FIFA 98 world cup site~\cite{arlitt2000workload} and are employed as the input traffic for \z~. We consider three different traces (TRi)--available in~\cite{WC98_Log}. The traces include clients web content requests from the web servers over the course of $24$ hours on three different days. The average number of requests per minute is $10,796$. However, the content requests are not uniformly distributed over time, and demonstrate the \emph{slashdot effect}, i.e.,~sudden and relatively temporary surges in traffic. As a result of the slashdot effect, the response time of the servers increases above the acceptable threshold and causes \late~~issues for the affected clients. The employed adaption policy $\pi$ addresses \late~~via \enlist~~or \reduce~~actions. The two actions however could cause \budget~,~\utiliz~, and~\quality~ issues which are dealt with once the slashdot effect wears off via~\discharge~and~\increase~. 
\newline\textbf{Deterministic hybrid planner}
We implemented a deterministic coordinating hybrid planner that uses predefined thresholds on quality attributes of interest\eg response time, as constraints. The proposed hybrid planner, \chp henceforth, does not support runtime adjustments of its control parameters and considers look-ahead, planning, and execution horizons with deterministic and predefined sizes and, as a result, exhibits smaller planning overhead at runtime.
\chp~takes current state $s$, set of available policies $\Pi$, current response time $RT_{curr}$, and response time threshold $RT_{thr}$ as inputs.  If $RT_{curr}$ exceeds $RT_{thr}$, \chp switches to a policy with a smaller $ET(plan)$, otherwise, a more time-intensive policy obtaining higher quality is employed. 
 %
%In the following we elaborate on the required information for \app as well as the setting to materialize its functionality at runtime using~\z~~from Section~\ref{sec:CaseStudy}. 
%

%We discuss if said concerns require \emph{meta level} control (i.e.,~\ml~) or can be dealt with at the primitive \emph{adaptation level} (i.e.,~\al~). 
%A SAS is intended to operate in highly dynamic environments with changing goals and requirements.

%In the context of our motivating example, during surges of client traffic in \z~~where more adaptation issues such as \late~~are likely to occur, employing $\pi_S$ that can provide timely rather than optimal adaptation plans is more desirable. However, once the traffic surge calms down, the utilization of the servers can be optimized via switching to $\pi_D$ that can provide optimal adaptation plans. 
	\vspace*{-1mm}
\subsection{Experiments}\label{subsec:experimnet}
\noindent\textbf{Policies}
The employed hybrid planners combine a static policy~($\pi_S$) and a dynamic policy~($\pi_D$) from~\cite{SG_TAAS} to equip the target system with self-adaptation capabilities. The policies conform to the definition of the static and dynamic adaptation policies in Section~\ref{subsec:RBSAS} respectively and use the utility function~$U_{znn}$ as their objective function.
$\pi_S$ uses design-time estimations for $EU$--see~Section~\ref{sec:CaseStudy}. Therefore, at each state $s$, for each applicable $a$, the expected effect on $EU$ is predetermined. We have shown in~\cite{SG_ICAC_17} that this policy is sub-optimal in terms of overall utility but fast in terms of adaptation time.  $\pi_D$ uses the IBM ILOG CPLEX constraint solver for selecting actions that optimize the utility. %Specifically, it uses the utility function $U_ {znn}$~in~(\ref{eq:Znnutility})  as its objective function. 
%The task of selecting the action $a$ for each each adaptation issue and ordering them for execution are defined as the optimization problem. 
We have shown in~\cite{SG_ICAC_17} that while this policy finds the optimal target state at each adaptation step, it can exhibit long planning times. Runtime conditions trigger a switch between $\pi_S$ and $\pi_D$. For example, during surges of client traffic in \z~~where more adaptation issues such as \late~~occur, employing $\pi_S$ that provides timely rather than optimal adaptation plans is beneficial. However, once the traffic surge calms down, the utilization of the servers can be optimized via switching to $\pi_D$. % that can provide optimal adaptation plans. 
The switch from policy $\pi_{D}$ to $\pi_{S}$ is charged with a cost $c_{DS}=2$ and  $c_{SD}=200$ applies to the switch in the opposite direction. The considered costs are independent of the runtime conditions and are defined relatively and based on the measurements of the policy deployment times. %Loading the constraint solver and activating $\pi_D$ is 100 times more expensive than activating $\pi_S$.
%$\pi_S$ uses design time estimations for decision making while $\pi_D$ uses the IBM ILOG CPLEX constraint solver for selecting actions that optimize the  utility function. 
%Each experiment is simulated for $24$ hours of input traces. 
\newline\textbf{Setting}~The experiments are repeated and averaged over $1000$ simulation runs.
The reported utility values are Normalized Accumulated Utility ($NAU$) with $ NAU=U_{znn}-c_{ij}$.
We executed \chp with $RT_{thr}=1~sec$. The same is set as the initial value to define the \emph{high} range for $RT_\mathcal{I}$ in the \app variants. $RT_{min}=0.1~sec$ is used as the initial value for the \emph{optimal range} and the $RT_\mathcal{I}$ values in between are considered as \emph{inRange}--see Algorithm~\ref{Algorithm:hypzn}. Note that the threshold values in the \app variants are not deterministic and may change at runtime. The size of the look-ahead, planning, and execution horizons in \chp is set to $\infty$, thus, \chp plans for \emph{all} the existing issues and executes the \emph{complete} plan.
 \begin{table}[]\centering
	\vspace*{-6mm}
	\caption {   NAU over $24$ Hours}
	\label{tab:AccUtility} 
	\vspace*{-2mm}
	\resizebox{7.5cm}{!}{
		\begin{tabular}{l|ccccc}\toprule
			Trace&\intr&\extr	& \chp&$\pi_D$&$\pi_S$\\
			\cmidrule{1-6}% \cmidrule{3-4}	
			
			TR1& 0.79&1&0.81&0.51&0.33\\
			TR2&0.55 &0.83&1&0.49&0.38\\
			TR3&0.76 &1&0.85&0.55&0.42\\
			
			\cmidrule{1-6}% \cmidrule{3-4}	
			
		\end{tabular}
	}
	\vspace*{-6mm}
\end{table}

\z~~is equipped with self-adaptive properties via adding a MAPE-K loop, henceforth \emph{adaptation loop}, to the system. \extr is realized in an additional loop that is added on top of the adaptation loop--see~Fig.~\ref{fig:MetaSelfAw}-(a). \intr implements the internal design for meta-self-awareness--see~Fig.~\ref{fig:MetaSelfAw}-(b).
The sampling interval $\mathcal{I}$ for \extr indicates that the meta-awareness loop in the \extr is executed once for every $\mathcal{I}$ executions of the adaptation loop.  In \intr however, $\mathcal{I}=1$ as the hybrid planner is embedded in the adaptation loop and  is executed at the same frequency as the adaptation loop--see Section~\ref{subsec:design}. 
\newline\textbf{Results}
Table~\ref{tab:AccUtility} shows $NAU$ during $24$ hours of the input traces for \z~. $\pi_D$ and $\pi_S$ exhibit basic planning for SAS and do not employ any hybrid planning, therefore, $c_{ij}=0$. \extr is executed with $\mathcal{I}=5$. Table~\ref{tab:ex} presents results of sensitivity analysis for  $NAU$ obtained by \extr with different execution intervals $\mathcal{I}$. Note that \extr with $\mathcal{I}=1$ is identical to \intr.

\reqone~\textbf{how do internal and external designs for meta-self-awareness affect \app?} As shown in Table~\ref{tab:AccUtility}, in two out of the three experiments, \extr achieves higher accumulated utility compared to \intr. The reason is that \extr has relatively larger sampling intervals\ie execution timescale, that provides an extended monitoring period. \intr holds a localized view of the system load that is limited to its relatively small monitoring period, i.e.,~since the last execution of the adaptation loop, and sets its control parameters and  switches the employed policies accordingly. The results in Table~\ref{tab:AccUtility} suggest that the relatively small execution timescale of the meta-awareness loop in \intr may lead to pre-mature decisions due to insufficient and localized information and, as a result, the hybrid planner is likely to demonstrate nervous and volatile behavior regarding policy switch.
\begin{table}[t]\centering
\vspace*{-6mm}
\caption {  NAU  of  \extr with Different $\mathcal{I}$ over $24$ Hours}
\label{tab:ex} 
\vspace*{-2mm}
\resizebox{9cm}{!}{
	\begin{tabular}{l|lcccccc}\toprule
		
	&$\mathcal{I}=1$ &$\mathcal{I}=2$ 	& $\mathcal{I}=3$ &$\mathcal{I}=4$ &$\mathcal{I}=5$ &$\mathcal{I}=10$ &$\mathcal{I}=15$ \\
		\cmidrule{1-8}% \cmidrule{3-4}	

		\z~~-~TR1& 0.79&0.81&0.85&0.92&1&0.43&0.25\\

		\cmidrule{1-8}% \cmidrule{3-4}	
		
	\end{tabular}
}
\vspace*{-6mm}
\end{table}
%
%In the experiment with Grid5000 trace for \mrubis \intr outperforms \extr. The reason is that the IAT of the adaptation issues in this trace is significantly smaller compared to the other employed input traces. Smaller IAT between adaptation issues demand more frequent planning in the adaptation loop. In this case, as discussed in Section~\ref{subsec:control}, more frequent control calculations are beneficial since the system and its context  change in high pace. Therefore, when SAS experiences adaptation issues with high arrival rate, smaller sampling intervals $\mathcal{I}$~for the controller are more beneficial since \app variants adjust their control parameters at each sampling interval.
%
The results in Table~\ref{tab:AccUtility} and~\ref{tab:ex} suggest that the utility of the hybrid planners is affected by the characteristics of the input traces. Moreover, as confirmed by Table~\ref{tab:ex}, larger values for execution intervals of \extr result in sub-optimal adaptations; \extr with $\mathcal{I}=15$ achieves only $25\%$  of the optimal utility for TR1.

\reqtwo~\textbf{how does \app perform in comparison to a deterministic hybrid planner?} 
Table~\ref{tab:AccUtility} shows that for TR2 in \z~, \chp obtains higher accumulated utility compared to the \app variants. Analysis of the TR2 characteristics revealed that the request arrival rate in TR2 is only 20\% of the one in TR1 and 36\% of TR3. The results together with the analysis of the input trace characteristics confirm the fact that runtime conditions\ie~characteristics of the input traces significantly affect the utility of the adaptation.
The results of Table~\ref{tab:AccUtility} show that in two out of the three experiments, \extr outperformed \intr and \chp. For the  trace with less extreme characteristics\ie~TR2, the predefined values of the control parameters and thresholds in \chp are beneficial and outperform the \app variants. Overall, results suggest that \extr suits best the volatile operation conditions\eg~input traces with more extreme characteristics.% in our experiments. 

%As discussed earlier and detailed in Table~\ref{tab:FailureLogs} Grid5000 trace includes relatively large bursts of adaptation issues that arrive in high pace. 

\begin{figure}[b]
%	\vspace*{-3mm}
	\begin{centering}
		\includegraphics[width=\linewidth]{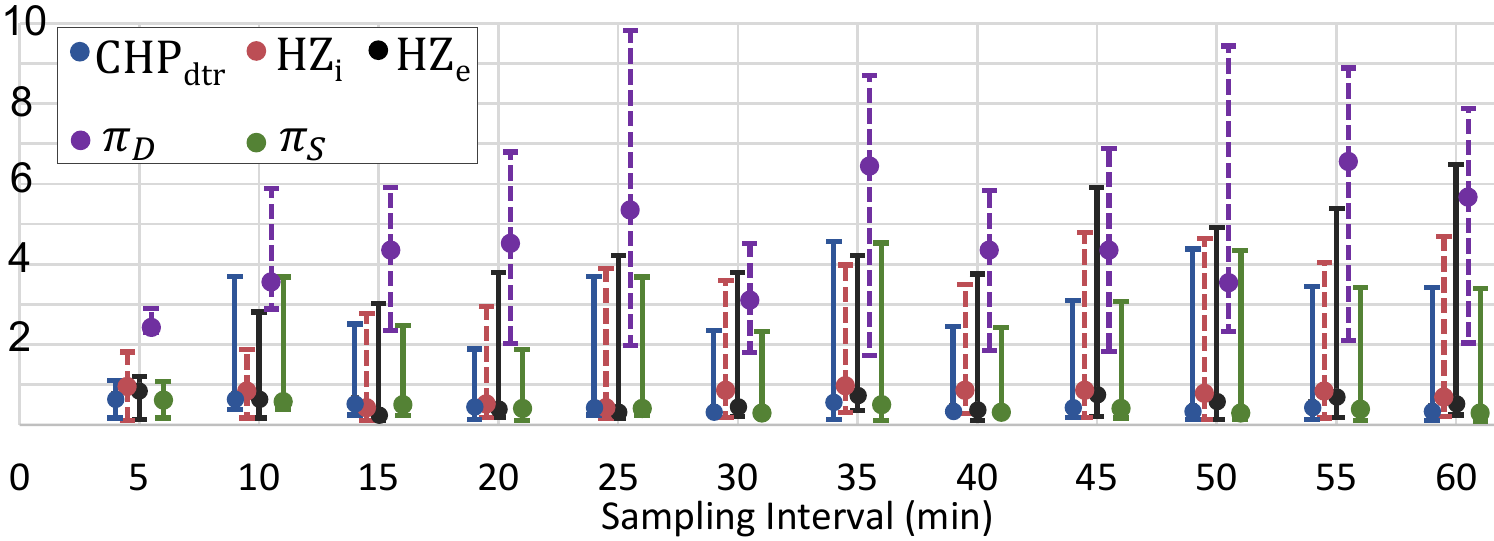}
		%	\vspace*{-2mm}
		\caption{Average-max-min request response time (sec) in \z~. }
		\vspace*{-6mm}
		\label{fig:RTZnn}
	\end{centering}
	\vspace*{-2mm}
\end{figure}

 Fig.~\ref{fig:RTZnn} presents average request response times for clients in \z~ over $60~min$. \extr is executed with $\mathcal{I}=5$. Each measurement at sampling interval $j$ shows the average response times (bullets)~as well as the maximum and minimum response times (vertical bars) during the last $5~min$. Note that the hybrid planning overhead also affects the response time values. Table~\ref{tab:znn60min} shows the corresponding $NAU$ of the planners at the end of $60~min$. \chp uses predefined and deterministic values for the control parameters, thus, compared to the \app variants, has a smaller planning overhead and exhibits smaller response times. In addition to the three hybrid planners, Fig.~\ref{fig:RTZnn} also includes response times for their constituting individual adaptation policies\ie~$\pi_D$ and $\pi_S$. $\pi_S$  has similar response times to \chp. This is due to the deterministic decision-making in \chp that as soon as response time raises above $1~sec$, \chp switches to the static policy for planning.
 Compared to \chp, \extr has slightly higher average response times. However, as shown in Table~\ref{tab:znn60min}, \extr obtains $35\%$ higher accumulated utility over $60~min$ in the same experiment. Response time values for \intr~are higher than \extr and \chp. Despite its higher response times, Table~\ref{tab:znn60min} shows that  \intr obtains $14\%$ higher accumulated utility compared to \chp. 
%	\vspace*{-1mm}
  \begin{table}[t]\centering
	\vspace*{-6mm}
	\caption {  NAU over $60$ min for  \z~}
	\label{tab:znn60min} 
	\vspace*{-2mm}
	\resizebox{6.5cm}{!}{
		\begin{tabular}{lcccc}\toprule
			\intr&\extr	& \chp&$\pi_D$&$\pi_S$\\
			\cmidrule{1-5}% \cmidrule{3-4}	

			0.79&1&0.65&0.59&0.51\\

			\cmidrule{1-5}% \cmidrule{3-4}	

		\end{tabular}
	}
	\vspace*{-6mm}
\end{table}

\reqthree~\textbf{what are the effects of  hybrid planning on the quality and timeliness of the adaptation?} The $NAU$ values  in Table~\ref{tab:AccUtility} suggest that employing hybrid planning improves the utility of a SAS compared to their individual constituting policies. 
 $\pi_D$ in~Fig.~\ref{fig:RTZnn} exhibits the highest response time, it also has significantly high \emph{maximum} response time values\ie~$10~sec$. In case of other planners, the maximum response time does not exceed $5~sec$. While results in~Fig.~\ref{fig:RTZnn} show relatively low response times for $\pi_S$, it obtains only 51\% of the optimal accumulated utility in~Table~\ref{tab:znn60min} .

The baseline solutions that employ a single planner are likely to exhibit sub-optimal behavior caused by the changing operation condition. We have shown in~\cite{SGHGcomputers2020} that the choice of the adaptation policy in a SAS should be steered with respect to the  characteristics of the input trace, otherwise, the employed policy may render sub-optimal at runtime. Thus, as also confirmed by our experiments, employing a hybrid planner, utilizing either deterministic or adjustable parameters, results in  improvements of the utility (Table~\ref{tab:AccUtility} and~\ref{tab:znn60min}) as well as the timeliness (Fig.~\ref{fig:RTZnn})~of the adaptation.

 %. Adding meta-awareness to SAS , that is, increased awareness of itself and its context, enables SAS to realize and consequently benefit from the increased awareness. Our experiments show that as a results of this increased awareness, both the \extr~~and the \intr~~variants realize the runtime capacities and use them to improve the system utility over time as opposed to the \bl~~variant that ignores the runtime capacities.
%FULL Version
%The results suggest that employing meta-awareness capabilities, either as External or Internal design, results in improvements in the system overall utility compared to the \bl~. Adding meta-awareness to SAS , that is, increased awareness of itself and its context, enables SAS to realize and consequently benefit from the increased awareness. Our experiments show that as a results of this increased awareness, both the \extr~~and the \intr~~variants realize the runtime capacities and use them to improve the system utility over time as opposed to the \bl~~variant that ignores the runtime capacities. 
%

  %the \extr~~variant does not react to the relatively small turbulences in the number of requests since it holds a longer monitoring period. 
  
\vspace*{-1mm}
%  \newline\textbf{Discussion on Meta-awareness for Hybrid planner } 
\subsection{Discussion }
There exist various ways to provide the hybrid planner in a SAS with history of the system past executions and observations to enable more informed decisions,~i.e.,~history-aware self-adaptation schemes, e.g.,~\cite{LSHistory}. Considering hybrid planning as a case for meta-awareness has the following advantages and limitation; the external design provides for a global view on the target system and the adaptation process. This allows for observing phenomena with global scope that are not observable at the awareness level. \extr supports explicit separation of concerns at the architecture level and allows for re-usability, easier maintenance, and  independent evolution of each level. Consequently, separate and independent mechanisms for observing, analyzing, and reasoning logic may be employed by each level.	
The internal design in \intr limits the controller in the meta-awareness subject to a localized view of its object. In this design, the meta-awareness logic is dispersed throughout the awareness level. The intertwined realization of the awareness and meta-awareness together with the embedded and dispersed meta-awareness logic makes it challenging to reason about the outcome of the meta-awareness, making composability and re-usability difficult to achieve.   

 	\vspace*{-2mm}
\section{Conclusion and Future Work}\label{sec:conclusion}
\noindent
In this paper, we presented \app, a solution for hybrid planning in SAS.  \app leverages receding horizon control to utilize runtime information for decision-making. We proposed two alternative designs conforming to the meta-self-aware architectures to engineer \app for SAS. %We showed that while different operation conditions of SAS affect the utility and  timeliness of the hybrid planners, it is beneficial to consider hybrid planning over the use of single adaptation policy.
We showed that hybrid planning, realized either as meta-awareness capabilities, or as a basic deterministic heuristic, is beneficial for SAS as it provides extended control flexibility at runtime. Our experiments suggest that, compared to the deterministic alternative,  hybrid planners that utilize runtime information to dynamically adjust their decision-making  are more beneficial in copping with the volatile operation conditions. Moreover, \app has the characteristics of a generic hybrid planner and considers the  adaptation policies as black-box and can coordinate arbitrary adaptation policies. %\eg~input traces with more extreme characteristics.
% in our experiments.  
 %a SAS with meta-awareness properties realizes runtime capacities that improve its behavior in multiple ways; Enhancements to the adaptation policy, adjustments of the control parameters, and devising new actions to better cope with the  operation conditions are examples of runtime capacities that a meta-self-aware SAS may realize. 
Investigating the concurrent execution of adaptation policies in \app is a subject of future work. Moreover, in addition to the functional aspects, we plan to study the architectural aspects of realizing hybrid planning as a meta-self-awareness property in SAS.

\vspace*{-2mm}

\bibliographystyle{abbrv}
\bibliography{Refs}

%\begin{thebibliography}{00}
%\bibitem{b1} TODO
%\end{thebibliography}

\end{document}